\pdfoutput=1

\documentclass{sig-alternate}

\usepackage[utf8]{inputenc}
\usepackage[T1]{fontenc}

\usepackage{multirow}
\usepackage{graphicx} 
\DeclareGraphicsExtensions{.pdf,.eps,.png,.jpg,.mps}
\usepackage{adjustbox}
\usepackage{listings}

\begin{document}
%
\conferenceinfo{COP '15}{July 05 2015, Prague, Czech Republic \\
Copyright is held by the owner/author(s). Publication rights licensed to ACM.}
\copyrightetc{ACM \the\acmcopyr}
\CopyrightYear{2015} 
\crdata{978-1-4503-3654-3/15/07 \ ...\$15.00. \\ 
DOI: http://dx.doi.org/10.1145/2786545.2786552}

\title{Towards a Decoupled Context-Oriented Programming Language for the Internet of Things}

\numberofauthors{6} 

\author{
\alignauthor
Baptiste Maingret\\
       \affaddr{University of Lyon}\\
       \affaddr{INSA-Lyon, Telecommunication Dpt}\\
       \affaddr{F-69621, Villeurbanne, France}\\
       \email{baptiste.maingret@insa-lyon.fr}
\alignauthor
Frédéric Le Mouël\\
       \affaddr{University of Lyon}\\
       \affaddr{INSA-Lyon, CITI-INRIA Lab}\\
       \affaddr{F-69621, Villeurbanne, France}\\
       \email{frederic.le-mouel@insa-lyon.fr}
\alignauthor 
Julien Ponge\\
       \affaddr{University of Lyon}\\
       \affaddr{INSA-Lyon, CITI-INRIA Lab}\\
       \affaddr{F-69621, Villeurbanne, France}\\
       \email{julien.ponge@insa-lyon.fr}
\and  
\alignauthor 
Nicolas Stouls\\
       \affaddr{University of Lyon}\\
       \affaddr{INSA-Lyon, CITI-INRIA Lab}\\
       \affaddr{F-69621, Villeurbanne, France}\\
       \email{nicolas.stouls@insa-lyon.fr}
\alignauthor 
Jian Cao\\
       \affaddr{Shanghai Jiao Tong University}\\
       \affaddr{800 Dongchuan Road}\\
       \affaddr{Shanghai, 200240, China}\\
       \email{cao-jian@sjtu.edu.cn}
\alignauthor 
Yannick Loiseau\\
       \affaddr{Blaise Pascal University}\\
       \affaddr{24 Avenue des Landais}\\
       \affaddr{F-63170 Aubière, France}\\
       \email{yannick.loiseau@univ-bpclermont.fr}
}

\maketitle
\begin{abstract}
Easily programming behaviors is one major issue of a large and reconfigurable deployment in the Internet of Things. Such kind of devices often requires to externalize part of their behavior such as the sensing, the data aggregation or the code offloading. Most existing context-oriented programming languages integrate in the same class or close layers the whole behavior. We propose to abstract and separate the context tracking from the decision process, and to use event-based handlers to interconnect them. We keep a very easy declarative and non-layered programming model. We illustrate by defining an extension to Golo - a JVM-based dynamic language.
\end{abstract}




\section{introduction}
Easily programming behaviors is one major issue of a large and reconfigurable deployment in the Internet of Things. Such kind of devices often requires to externalize part of their behavior such as the sensing, the data aggregation or the code offloading. Most existing context-oriented programming languages integrate in the same class or close layers the whole behavior. 

In this article, we propose 
\begin{itemize}
\item to abstract and separate the context tracking from the decision process. Indeed, most context-oriented languages use event-condition-action rules embedded in the business class or in the context definition. This definition is rather restrictive and would benefit from more complex decision-making mechanism such as neural network or machine learning;
\item to externalize these interactions by using API-defined context and decision maker, and interconnecting them with event-based handler mechanisms;
\item to demonstrate the feasibility of such externalization by developing Congolo - \emph{Contextual Golo}, an extension to Golo - a JVM-based dynamic language \cite{ponge_golo_2013}. Preliminary performances are given.
\end{itemize}

Section \ref{related-work} details related works. Section \ref{architecture} introduces Congolo: language enhancements and architecture. Section \ref{implementation} goes into Congolo implementation, and finally, section \ref{results} gives preliminary results. 

\section{Related Work}
\label{related-work}

Several COP languages exist based on a variety of programming languages. Most of them implements the context with the use of layers. Layers include context tracker mechanisms and context reaction ones. In order to compare these languages, we can thus focus at first on the way the context are defined, then how they are activated and finally how they are used throughout the code. Table \ref{table:coplanguages} proposes a summary.

  \begin{table*}
  \begin{adjustbox}{center}
  \begin{tabular}{l p{3cm} c c c c c c c c c c c}

  \multicolumn{2}{l}{Languages} & \rotatebox{90}{ContextJ} & \rotatebox{90}{JCop} & \rotatebox{90}{EventCJ} & \rotatebox{90}{NextEJ} & \rotatebox{90}{ContextLua} & \rotatebox{90}{ContextErlang} & \rotatebox{90}{EventJava} & \rotatebox{90}{ServalCJ} & \rotatebox{90}{ECaesarJ} & \rotatebox{90}{Subjective-C} \\

  \hline
  \multirow{2}{*}{Context declaration}
    & Layer &   X & X & X & - & X & X & - & - & - & - \\
    & Class &   - & - & - & X & - & - & X & X & X & X \\
    
  \hline
  \multirow{2}{*}{Context type}
    & State &   X & X & X & X & X & X & - & X & X & X \\
    & Data  &   - & - & - & - & - & - & X & - & - & - \\    

  \hline  
  \multirow{2}{*}{Layer declaration}
    & Layer-in-class &   X & X & X & X & X & X & X & X & X & X \\
    & Class-in-layer &   - & - & - & - & - & X & - & - & - & - \\

  \hline
  \multirow{2}{*}{Layer implementation}
    & Class &       - & - & - & - & - & - & - & - & X & - \\
    & Layer type &  X & X & X & X & X & - & - & X & - & X \\    

  \hline  
  \multirow{3}{*}{Scope}
    & Instance &        - & - & X & - & X & - & X & X & X & X \\
    & Thread &          X & X & - & - & - & - & X & X & - & X \\
    & Global &          - & - & - & - & - & X & X & X & - & - \\
  \hline
  \multirow{2}{*}{Active context tracking}
    & Push to layer stack &             X & X & X & - & - & X & X & X & - & X \\
    & Directly change method lookup &   - & - & - & - & - & - & X & - & - & X \\

  \hline    
  \multirow{3}{*}{Implementation}
    & Language extension &   X & X & - & - & X & X & X & - & X & X \\
    & New language &         - & - & X & X & - & - & - & X & - & - \\

  \end{tabular}
  \end{adjustbox}
  \caption{COP languages comparison}
  \label{table:coplanguages}

  \end{table*}

In many languages, context is integrated directly in the business code by the means of layers. It is the case in JCop \cite{appeltauer_declarative_2013}, ContextJ \cite{appeltauer_improving_2009}, ContextErlang \cite{ghezzi_context_2010}, ContextLua \cite{wasty_contextlua:_2010} or EventCJ \cite{kamina_eventcj:_2011}. In ECaesarJ \cite{nunez_declarative_2009}, NextEJ \cite{kamina_towards_2009}, ServalCJ \cite{Kamina:2015:GLA:2724525.2724570} or Subjective-C \cite{RELEASeD-2010-857551}, the context is declared separately for instance by class inheritance or \lstinline|context| tag declaration. In EventJava \cite{jayaram_context-oriented_2009}, the context is considered to be an event with specific data and is defined alongside the event consumer. Thus, when the event is triggered, the context is embedded into it.

The activation of the context, or layers depending on the language, often use the keyword \lstinline|with| \cite{haupt_contextj:_2011} \cite{appeltauer_declarative_2013} \cite{kamina_towards_2009} \cite{wasty_contextlua:_2010}. Instead ContextErlang \cite{ghezzi_context_2010}, ECaesarJ \cite{nunez_declarative_2009} or EventCJ \cite{kamina_eventcj:_2011} use a different approach where contexts are activated independently of the running program by the mean of events. In ECaesarJ \cite{nunez_declarative_2009}, the activation is done by the use of specific events declared in the context object which can be triggered by others events, by method invocation or by composite expression. In EventCJ \cite{kamina_eventcj:_2011}, they declare transition rules based on events that will activate or deactivate contexts. In ServalCJ \cite{kamina_unified_2013} \cite{ Kamina:2015:GLA:2724525.2724570}, where the distinction is made between context and layers, the context state is changed by specific actions and thus can be activate in the program, whereas layers are activated according to the state of one or multiple contexts and thus are activated implicitly. Finally in EventJava \cite{jayaram_context-oriented_2009} where the context is defined by specific values of information when the event is triggered, the activation of the context is global but its particular definition, i.e. the values of the information that it holds, are defined by events.

Finally, we need to consider how the program takes into account the different contexts to adapt and modify the execution. Two strategies are often found when the language use a layer paradigm: layer-in-class and class-in-layer implementation. In the first one, the layers and corresponding behaviors are implemented inside the class such as in \cite{appeltauer_improving_2009} \cite{appeltauer_declarative_2013} \cite{wasty_contextlua:_2010} or as in \cite{kamina_eventcj:_2011} \cite{kamina_unified_2013} \cite{Kamina:2015:GLA:2724525.2724570}. NextEJ \cite{kamina_towards_2009} follows the same approach but differs by the fact that they use roles defined in the context classes to implement the different behaviours. ContextErlang \cite{ghezzi_context_2010} states that both methods have their advantages and thus offers the possibility of using the both of them. ECaesarJ \cite{nunez_declarative_2009} offers two distinct ways of using the current context. First, it is possible to bind specific events to the events triggered by the context changes or to directly query the context state with the help of the method \lstinline|isActive()| of the context. In EventJava \cite{jayaram_context-oriented_2009}, as the context is represented by variables bound to an events, the context-dependent behaviors are implemented using this data.

One can encounter different type of implementation for COP languages. The easiest one might be to develop an API such as ContextJ \cite{appeltauer_improving_2009}. One can also extend existing languages with new keywords such as in \cite{clarke_semantics_2009}, ContextJ \cite{haupt_contextj:_2011}, JCop \cite{appeltauer_declarative_2013}, NextEJ \cite{kamina_towards_2009}, ContextErlang \cite{ghezzi_context_2010} or Subjective-C \cite{RELEASeD-2010-857551}. Finally some approaches propose a new COP language as EventCJ \cite{kamina_eventcj:_2011} or ServalCJ \cite{kamina_unified_2013} \cite{Kamina:2015:GLA:2724525.2724570}.

\begin{center}
\begin{figure}[!b]
\centering
\includegraphics[width=0.9\columnwidth, bb=0 0 619 325]{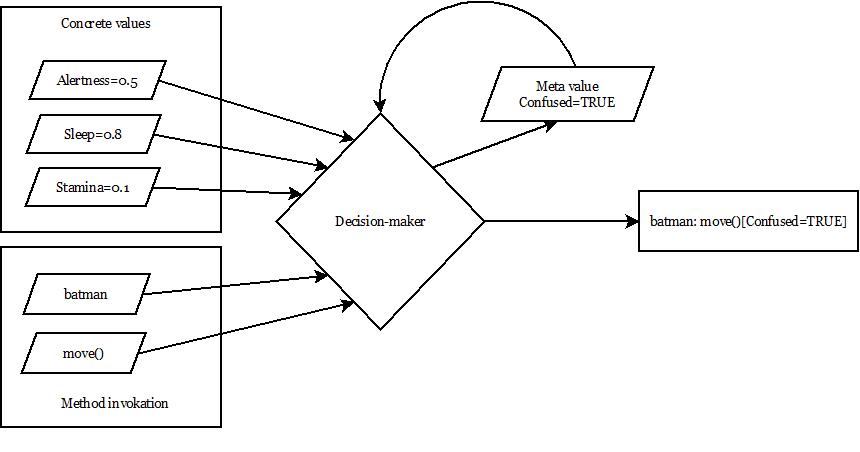}
\caption{Concrete and meta values in context representation}
\label{figure:concretemetavalues}
\end{figure}
\end{center}

\section{Decoupled Architecture}
\label{architecture}

We present in this section the Congolo language enhancements and the interpreter architecture.

\subsection{Two-level Context Management}

The declaration of a context is done using the keyword \lstinline|context| as shown in Listing \ref{listing:congolocontext}. The following expression must be a tuple composed of class instances implementing a specific \lstinline!Context! interface. Scope of this context is the module where it is declared.

\begin{lstlisting}[float, language=Java, caption=Congolo context example, label={listing:congolocontext}]
contexts  =  [ConfusedHero(), Weather()]
\end{lstlisting}

Where most approaches handle context as a simple state system with variables - activated or deactivated, we decided to allow a more flexible way of handling the context by introducing two levels of context values: meta values and concrete values. The meta values are used into the business code to declare the different layers. Concrete values are used inside the decision-maker to compute and output the meta values and thus to decide which layers to activate, as shown in Figure \ref{figure:concretemetavalues}. Hence, the context is not directly used to invoke methods.

\subsection{Function-grained Layers}

\label{subsection:layers}
Layers are usually defined as alternative methods to base methods (that are context independent) that will eventually be triggered depending on the context. The declaration is done with the help of a keyword such as \lstinline|layer|, allowing to define multiple methods belonging to one layer. In our case, we propose an alternate method definition to indicate layers by using the at symbol and appending them to the method's parameter list: \lstinline!function myFunc = |arg1|@(contextA=METAVALUE1)! as in Listing \ref{listing:congololayers}. This choice was made because of the compatibility of this definition with the different function definitions allowed by Golo and because we think it looks like it really belongs to the language rather than having been added in an ad-hoc manner.

In other languages, it is usually possible to call the base method from the layered one. We propose this in a similar manner by using the keyword \lstinline|proceed()| that will refer to the base method from the layered one. In addition we propose syntactic support for two basic cases where the developer wish to call the base method before or after the layer as shown in Listing \ref{listing:congololayers}.

When multiple layers are activated at the same time, the usual proposed way is to simply take them in a LIFO manner as if they were stacked. However in our case the method to be called is chosen by the decision-maker. Thus in the case were multiple contexts, i.e. multiple meta values, are activated at the same time, again the system will make the decision to chose which methods are to be called.

\begin{lstlisting}[float, language=Java, caption=Congolo layers example, label={listing:congololayers}]
function Hero = || {
  return DynamicObject():
    contexts([ConfusedHero, Weather]):
    # base function
    define("getPosition", |this| -> ...):
    define("move", |this, dir| -> ...):
    define("getPos"; |this| -> ...):
    # layered function, invoked before base method
    define("getPos", |this, direction|@(ConfusedHero=TRUE)+ -> ...):
    # layered function, invoked after base method
    define("move", |this, dir|+@(ConfusedHero=TRUE) -> ...):
    # layered function, with a call to the base function from within the layer
    define("move", |this, dir|@(ConfusedHero=TRUE) = {
      ...
      proceed(dir)
      ...
    }
}
\end{lstlisting}

\subsection{Decision-Maker}

One singular aspect of our proposition regarding previous ones is the way layers are activated. In many approaches, it was usually done in an ECA manner: when a specific context is activated and that a layered method is invoked, the corresponding layer is used. In that case, the decision is made at the programming step. It is of course possible to dynamically change the context at runtime but its binding to the layers is still static. In our case, we propose to delegate this decision to an external system, the decision-maker, which could be static just as in existing approaches but could also be based on more advanced algorithms and techniques from machine learning such as neural networks or genetic algorithms.  In this case, the system could actively learn during the runtime of the application. We offer two ways of defining which decision-maker is to be used to resolve the method to invoke. First it is possible to have a specific decision-maker per object, and declaring it by defining a \lstinline|decisionmaker| attributes in a dynamic object using a reference to an instance of a class implementing the Java interface \lstinline|fr.insa.lyon.congolo.api.DecisionMaker| as in Listing \ref{listing:congolodecisionmaker1}. If no \lstinline|decisionmaker| is defined in a object, all function invocation will refer to a global decision-maker instantiated at the program start.

\begin{lstlisting}[float, language=Java, caption=Congolo decision maker definition, label={listing:congolodecisionmaker1}]
function Hero = |decisionMaker| {
  return DynamicObject():
  decisionmaker(decisionMaker)
}

let misterPresident = my.app.myDecisionMaker()
let batman = Hero(misterPresident)
\end{lstlisting}

\begin{lstlisting}[float, language=Java, caption=Congolo example, label={listing:congolohero}]
let misterPresident = my.app.myDecisionMaker()
misterPresident: global(true)

let batman = Hero() # Context-dependent object
batman: move()        # Call to a layered method that will be handled by the local or global decision-maker

\end{lstlisting}

In order to further separate the decision-making process from the rest of the application, we propose that the decision-makers use events as primary inputs and outputs. This events can either contain data, such as data from the concrete values, object and method for method invocation, or context information (meta values). In addition, the output can either be a method invocation, corresponding to the decision made by the system, or an event that could be further used as an input for the system.

\section{From Golo to ConGolo}
\label{implementation}

In this section, we detail the implementation of the proof-of-concept we developed. It comes with a limited set of functionality and does not follow exactly the requirements of the contributions in terms of syntax or integration but the main aspect which is the separation between the decision-making system and the rest of the code is working and it is still sufficient to develop a simple use case and perform first experiments. 

\subsection{Context support in the language}
To provide the support for layered methods, we first modified the grammar of the language to introduce the corresponding elements as shown in Listing \ref{listing:implgrammar}, and then modified the AST representation \lstinline|fr.insalyon.citi.golo.compiler.parser.ASTFunction| to support the added context information. The grammar is defined using JJtree, a preprocessor for JavaCC.

\begin{lstlisting}[float, language=Java, caption=Grammar modification - Golo.jjt, label={listing:implgrammar}]
void Function():
{
  List<String> arguments = null;
  Token varargsToken = null;
  boolean compactForm = false;
  List<String> contexts = null;
  boolean contextual = true;
}
{
  ("|" arguments=Arguments() (varargsToken="...")? "|")?
  ("@(" contexts=Contexts() ")")?
  ...
}
\end{lstlisting}

To support multiple function definition in a same module, we used a renaming convention so that a function with a context information would be renamed as \lstinline|originalFunctionName_ _$context$__context|.  Golo uses an intermediate representation (IR) to ease the compiling process and we implemented this part at the translation from the ASTtree to this IR representation in the \lstinline|fr.insalyon.citi.golo.compiler.ParseTreeToGoloIrVisitor| class.

\begin{lstlisting}[float, language=Java, caption=IR modification - ParseTreeToGoloIrVisitor.java, label={listing:implastfunction}]
public Object visit(ASTFunctionDeclaration node, Object data) {
  ...
  if ((child instanceof ASTFunction) && (((ASTFunction) child).isContextual())) {
    nodeName = node.getName() + "__$context$__" + ((ASTFunction) child).getContexts().get(0);
  }
  ...
}
\end{lstlisting}

\subsubsection{From function call to the decision system}

\begin{center}
\begin{figure}
\centering
\includegraphics[width=0.9\columnwidth, bb=0 0 658 408]{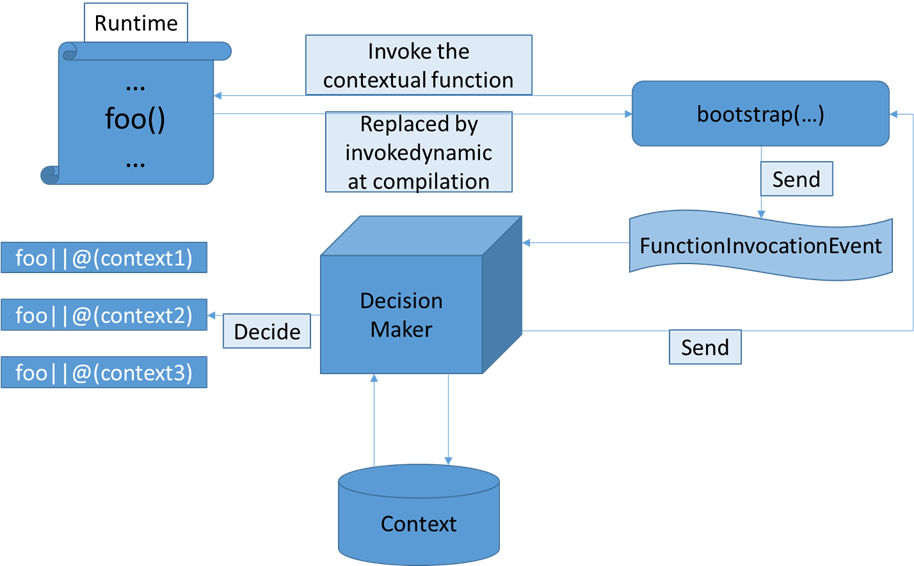}
\caption{Congolo invocation process}
\label{figure:congoloinvoke}
\end{figure}
\end{center}

In order to abstract and decouple the decision-maker from the rest of the code, we opted for an event-oriented system, as in Figure \ref{figure:congoloinvoke}. But to be able to use we first needed to add the support for contextual function invocation to Golo so that we could trick it as we wanted afterwards. We implemented this in the compilation process of Golo, whereas the decision-maker is designed as a Java API.

\subsubsection{Custom boostrap method}
First to be able to call the contextual function by their original given name, we changed the function invocation of contextual functions. Golo is based on the invokedynamic opcode to bind at the runtime the callsite to the correct target, as defined by the language. For this process it provides several utility classes in the \lstinline|fr.insalyon.citi.golo.runtime| package that acts accordingly to their name; FunctionCallSupport, MethodInvocationSupport, etc. We thus extended this package with the new class \lstinline|ContextualFunction CallSupport|. The Golo compiler replaces function calls, in the most generic terms of that, by invokedynamic instruction that are bound to specific bootstrap utility class according to their type (method, function, closure). In order to be able to use our newly designed bootstrap class we thus need to identify which of the function calls correspond to contextual function calls. We make this identification in the \lstinline|LocalReferenceAssignmentAndVerificationVisitor| class. We already know which functions are contextual so we build a list containing those so that we can latter check whether the name of the invoked function matches the name of a contextual function and marking it as such. Then in the bytecode generation process we simply use our custom bootstrap class for those function calls.

The contextual bootstrap method then wraps everything into a message that will be sent to the decision maker. Since the call needs to be synchronized (the bootstrap method should return a method handle that will later be used at the original callsite, we then block the current thread waiting for the answer from the decision-maker. The bootstrap class actually reacts on a specific type of message that triggers a function that will unlock the thread and finally configure the method handle according to the decision made by the decision-maker. Once the bootstrap method is made the call site is bound to the correct target as for the decision-maker.

\subsubsection{Event messaging}
The messaging system used is a simple hierarchical- namespace messaging system developped especially for Golo \cite{flemouel_gololang-messaging_2013}. These events are used to make the link between the function call and the decision-maker. The implementation is multi-threaded and the hierarchical design allows for interesting behaviors for a COP languages. It will be indeed easy to reach multiple decision-makers, or implement various scopes for the contexts.

The framework either uses a callback mechanism where you can register a specific handle to be invoked upon the reception of the message, or an interface where you can register an instance of a class implementing the \lstinline|gololang.concurrent.messaging.MessagingFunction|.

\subsubsection{The decision maker}
The decision-making part is designed as a Java API that defines a common \lstinline|DecisionMaker| interface. We also provided a \lstinline|DefaultDecisionMaker| class that implements some basic needed functionality allowing to quickly test our solution. The interface requires the implementation functions such as \lstinline|train| or \lstinline|init| but which are not used in our implementation because of its simplicity. They would be used in more advance systems such as a neural network. 

\subsubsection{Context management}
In order to be able to retrieve the adequate context regarding the function call, we register context objects with the help of a context manager that keeps track of which contexts were declared in which module as it is for now the only level we support for context declaration. When the decision-maker wants to retrieve the context information it queries the context manager with the module name and gets the list of the context objects declared for this module.

\section{Preliminary results}
\label{results}

The evaluation of programming languages can be difficult. Benchmarks are a common way of doing this, but there usually are pitfalls because of the numerous options of compilation or runtime, the type of hardware, the versions of software and so on. In COP languages, a common benchmark is to compare calls to layered functions versus calls to standard functions and see the overhead such as described in \cite{appeltauer_comparison_2009} where multiple languages were tested to create a benchmark. To do so they compare 10 methods layered to plain implementation. 

\subsection{Micro-benchmark performance}
Benchmarking is not an easy task especially for runtime-evaluated language. In addition they often do no reflect the future use or condition of the tested language. However it is still important to have an idea of how it performs relatively to others. In this perspective we designed a simple micro-benchmark to test specifically the predominant contribution of this paper which is the separation of the desision-making process with the rest of the code.

For that purpose we used the Java JMH tool developped by OpenJDK. This framework provides a harness for writing test to avoid common pitfalls \cite{jponge_benchmarking_2014}. It is organized as a Maven project where thanks to the JMH each test is run in a separate VM and is first run for warmups and afterwards for the measurement, so that the measures are consistent. The test were done on Windows 7 64 bits SP1 laptop Core 2 Duo @ 2.53GHz with 4 GB of RAM, with the build 1.8.0\_05-b13 of the Java Runtime Environment and the build 25.5-b02 of Hotspot 64 bits.

First we tested a call to a standard Golo module-level defined function versus a contextual one. Afterwards we tested the imbrication of 10 layered methods as proposed before.

The results shown in Table \ref{table:benchmarking} show a major loss in performance from Golo to Congolo. We did not indicate the exact error (less than 10 percents in each case) because the results were significant enough. This can be explained quite simply by the implementation of the solution. 

\begin{table}
\begin{adjustbox}{center}
\begin{tabular}{l p{2cm} | c | c | c | }
        \hline
                Test & Language & Score (nops/ms) \\
                
        \hline
        \multirow{2}{*}{Single invocation} 
                & Congolo & 29 \\
                & Golo & 6135 \\

        \hline
    \multirow{2}{*}{Layered invocation (10)} 
                & Congolo & 1.9 \\
                & Golo & 3998\\                 
\end{tabular}
\end{adjustbox}
        \caption{Benchmark results}
        \label{table:benchmarking}
\end{table}

First the invokedynamic instruction can be bound to different types of callsite: static, mutable and volatile. In the static case, once the bootstrap method is made there is no possible change of target thus the method can be directly called and the the link is thus optimized at runtime. In the other cases, and especially for mutable callsites (volatile is not use in Golo nor Congolo), in the case we want to check if we should change it we can guard the callsite with a test. This test will be checked before invoking the target and if it fails the bootstrapping will occur again. For instance, in the case of closures in Golo, each callsite has a cache which allows to check whether the target has changed or not and thus improving the overall performances. In our case, the test would need to check whether the context has changed or not since last time, test that is not done at this time and instead return false so that the bootstrap occurs all over again. Thus for each call, the system need to make the decision based on the context which explain in part the results.

Another point comes from the method use to communicate between the bootstrap method and the decision-maker. We opted for events so allow for more separation and because of its flexibility. However this was not done considering performances. The messaging environment is a single-thread (it does support multiple threads for a general purpose use, since it is basically and extension of the Java \lstinline|Executor|, however in our case we do not support multiple thread and concurrency in the design of Congolo, and in heavy stress, as it is the case in the benchmarking, it can lead to errors (deadlocks), which can be a bottle-neck. We check the difference between the messaging environment and direct calls to the functions. The results are shown in Table \ref{table:benchmarking_direct}. We can see that if the difference for Golo with previous results are in the range of the score error, it is not the case for Congolo where the results are significantly different and show an improvement with the previous implementation.

On overall Congolo offers quite poor performances compared to Golo. However it is to be noted that achieving high performance was not the main axis of development, and achieving from 2 or 3 function calls - for highly layered calls - to 30 function calls per millisecond appears to be enough for Internet of Things environments - as context changes are usually linked to quite stable physical phenomena \cite{RELEASeD-2010-857551}.

\begin{table}
\begin{adjustbox}{center}
\begin{tabular}{l p{2cm} | c | c | c |}
        \hline
                Test & Language & Score (nops/ms) \\
                
        \hline
        \multirow{2}{*}{Single invocation} 
                & Congolo & 55 \\
                & Golo & 5939 \\

        \hline
    \multirow{2}{*}{Layered invocation (10)} 
                & Congolo & 2.7 \\
                & Golo & 3840 \\                
\end{tabular}
\end{adjustbox}
        \caption{Benchmark results with direct calls to the Decision-maker}
        \label{table:benchmarking_direct}
\end{table}

\section{Conclusion and Future Works}

We presented Congolo \cite{bmaingret_congolo_2014}, a context-oriented language based on the Golo language, a dynamic JVM-based language. The main aspect and originality was to separate the reasoning part from the application code which was achieved by the use of an event-based messaging system and the invokedynamic instruction. Even if achieving 2 to 30 function calls per milliseconds is enough for Internet of Things environments, major performance improvements are required. The event-based messaging performances are encouraging since only slowing a function call from few milliseconds and can still be improved by multi-thread dispatching. The decision-maker however is slowing 100 times a function call, and branching decision and caching are part of future works. Achieving high-level context composition is also part of future work \cite{vallejos2010}, and will probably require specific optimizations regarding the performances.

\bibliographystyle{abbrv}
\bibliography{cop-ecoop-congolo15}

\end{document}